\documentclass[aps,twocolumn,superscriptaddress,showpacs,fixfloat]{revtex4}
\usepackage{amsmath}
\usepackage{graphicx}

\begin{document}
\bibliographystyle{apsrev}

\title{Orbital Effects of
In-Plane Magnetic Fields Probed by \\
Mesoscopic Conductance Fluctuations}

\author{D. M. Zumb\"uhl}
\affiliation{Department of Physics, Harvard University, Cambridge,
Massachusetts 02138}

\author{J. B. Miller}
\affiliation{Department of Physics, Harvard University, Cambridge,
Massachusetts 02138}
\affiliation{Division of Engineering and Applied Sciences, Harvard
University, Cambridge,
Massachusetts 02138}

\author{C. M. Marcus}
\affiliation{Department of Physics, Harvard University, Cambridge,
Massachusetts 02138}
\author{V. I. Fal'ko}
\affiliation{Physics Department, Lancaster University, LA1 4YB
Lancaster, United Kingdom}
\author{T. Jungwirth}
\affiliation{Institute of Physics  ASCR, Cukrovarnick\'a 10, 162 53 Praha
6, Czech Republic } \affiliation{University  of  Texas at  Austin, Physics
Department,  1 University   Station  C1600,   Austin  TX 78712-0264}
\author{ J. S. Harris, Jr.}
\affiliation{Departement of Electrical Engineering, Stanford
University, Stanford, California 94305}


\begin{abstract}
We use the high sensitivity to magnetic flux of mesoscopic conductance
fluctuations in large quantum dots to investigate changes in the
two-dimensional electron dispersion caused by an in-plane magnetic field.
In particular, changes in effective mass and the breaking of momentum
reversal symmetry in the electron dispersion are extracted quantitatively
from correlations of conductance fluctuations.  New theory is presented,
and good agreement between theory and experiment is found.
\end{abstract}

\maketitle

A simplified view of transport in a planar two-dimensional conductor, as
formed for instance by a semiconductor heterostructure, suggests that when
only the lowest quantized subband is occupied, an in-plane magnetic field
couples only to the electron spin, allowing the influence of an applied
magnetic field to be separated into spin and orbital parts. However, the
emerging picture of quantum transport in parallel fields \cite{Meyer,
Falko, Zumbuhl} has turned out to be surprisingly rich, indicating that
even modest parallel (i.e., in-plane) fields can have significant orbital
coupling, break time-reversal symmetry, and generate mesoscopic
conductance fluctuations with field-dependent correlations---even without
spin-orbit coupling or occupation of higher subbands.

In this Letter, we use the high sensitivity of mesoscopic conductance
fluctuations (CF's) to magnetic flux and time-reversal symmetry (TRS) to
examine in detail the orbital effects of an in-plane magnetic field,
$B_{\parallel}$, in a quasiballistic quantum dot formed in a GaAs/AlGaAs
2D electron gas (2DEG). Quantitative comparison of experiment and theory
developed here allows the effects of $B_\parallel$ on the electron
dispersion in a planar 2DEG, including an anisotropic effective mass and a
breaking of TRS (in spatially asymmetric confinement potentials), to be
distinguished using various correlation functions of CF's. Effects of
nonplanarity of the 2DEG are also included in the theory, and have
distinguishable signatures in the CF correlations. The significance of the
present work is to demonstrate experimentally that the effects of an
in-plane field go far beyond Zeeman coupling, but {\em cannot} be
characterized in terms of simple flux threading through the finite
thickness of a 2D electron layer. Also, this study shows that phase
coherent CF's can be used as a sensitive quantitative tool, much as one
uses a superconducting quantum interference device (SQUID).

\begin{figure}[b]
              \label{fig1ucfcorr}
\includegraphics[width=3.0in]{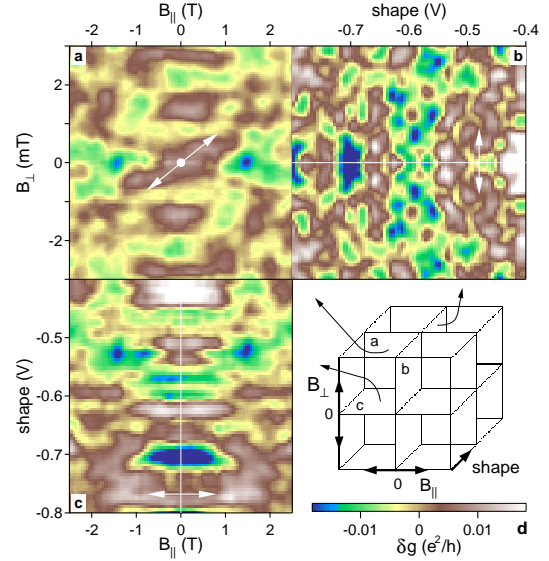}
        \caption{\footnotesize {Conductance fluctuations $\delta
        g(B_\parallel, B_\bot, V)=g(B_\parallel, B_\bot, V)-\langle
        g(B_\parallel, B_\bot, V) \rangle_V$ (color scale) through an
        $8\, \mathrm{\mu m^2}$ dot with one fully transmitting channel
        in each lead measured at $0.3$K with a) shape-gate voltage
        $V=-515\, \mathrm{mV}$, b) $B_\parallel=0$ and c) $B_\bot=0$,
        obtained from 2D slices of the three dimensional
     CF pattern, as indicated in d). }}
\end{figure}

Effects of parallel fields on quantum transport have been investigated in
2D systems, including metal films \cite{Metals}, silicon MOSFET's
\cite{SiMOSFET}, and GaAs/AlGaAs 2DEG's \cite{2DEG, 2DEGaniso}, as well as
in ballistic focusing geometries \cite{Focus}.  Those results were mostly
interpreted in terms of flux threading due to finite thickness in the
confined direction, surface roughness, or deformation of the Fermi circle
due to the field. Subband depopulation, decoupling of bilayer systems and
diamagnetic shifts caused by $B_\parallel$ have also been observed using
cyclotron resonance techniques \cite{Schlesinger}, magnetoresistance
measurements \cite{MR}, and tunneling \cite{Choi}, as well as optical
spectroscopy \cite{Reynolds}.  Related investigations based on quantum-dot
weak localization \cite{Zumbuhl} were only sensitive to the breaking of
TRS by a parallel field. Here, by using the full CF correlations, the
effect of $B_\parallel$ on the full electron dispersion is investigated,
and the various contributions are distinguished.

Two quantum dots, with areas $A=8\, \mathrm{\mu m^2}$ and $3\, \mathrm{\mu
m^2}$, made on the same wafer, were measured and showed similar behavior.
Data from the $8\, \mathrm{\mu m^2}$ dot (see Fig.~3, inset) will be
presented in detail. The dots are formed by lateral Cr-Au depletion gates
defined by electron-beam lithography on the surface of GaAs/AlGaAs
heterostructures grown in the [001] direction. The 2DEG interface is
$900\, \mathrm{\AA}$ below the wafer surface, comprising a $100\,
\mathrm{\AA}$ GaAs cap layer and a $800\, \mathrm{\AA\,
Al_{0.34}Ga_{0.66}As}$ layer with a $400\, \mathrm{\AA}$ Si doped layer
set back $400\, \mathrm{\AA}$ from the 2DEG. An electron density of $n=
2\times 10^{15}\, \mathrm{m^{-2}}$ and bulk mobility $\mu \sim14\,
\mathrm{m^2/Vs}$ (cooled in the dark) gives a transport mean free path
$\ell_{e}\sim1\, \mathrm{\mu m}$. Note that the $8\, \mathrm{\mu m^2}$ dot
contains of order $10^4$ electrons.

Measurements were made in a $ ^3$He cryostat at $0.3\, \mathrm{K}$ using
current bias of $1 \, \mathrm{nA}$ at $338\,\mathrm{Hz}$. Shape-distorting
gates were used to obtain ensembles of statistically independent
conductance measurements \cite{MarcusUCF} while the point contacts were
actively held at one fully transmitting mode each ($N=2$). In order to
apply tesla-scale $B_\parallel$ while maintaining sub-gauss control of
$B_\perp$, we mount the sample with the 2DEG aligned to the axis of the
primary solenoid (accurate to $\sim 1^\circ$) and use an independent
split-coil magnet attached to the cryostat to provide $B_\perp$
\cite{Folk}. The Hall effect measured in a separate GaAs Hall bar mounted
next to the quantum dot, as well as the location of weak localization
minima in transport through the dot itself (visible $B_\parallel \lesssim
2 T$) were used to determine the offset in $B_\bot$ (i.e. the residual
sample tilt), which was then corrected by computer control of the two
independent magnets.

The raw data consist of measured dot conductance
$g(B_\parallel, B_\perp, V)$ as a function of shape-distorting gate
voltage $V$ (inner loop of multiparameter sweeps), $B_\bot$, and
$B_\parallel$ (outer loop, swept from $-2.5\, \mathrm{T}$ to $+4\,
\mathrm{T}$ over $\sim20\, \mathrm{h}$), giving 20 independent shape, 15
independent $B_\bot$ and about 10 independent $B_\parallel$ samples.
Conductance fluctuations are found by subtracting the gate-voltage
averaged conductance over the measured range: $\delta g(B_\parallel,
B_\bot, V)=g(B_\parallel, B_\bot, V)-\langle g(B_\parallel, B_\bot, V)
\rangle_V$.

Figure 1 shows 2D slices of conductance fluctuations in the full 3D space
of $B_\parallel$, $B_\bot$, and $V$. Note that because
gate-voltage-averaged conductance is subtracted from the fluctuations,
weak localization effects on $\langle g\rangle$ are not evident in Fig.~1.
A principal result is already evident in Fig.~1: The horizontally
elongated features around $|B_\parallel|\sim0$ in Fig.~1(c) show
qualitatively that CF's are less sensitive to $B_\parallel$ in the
vicinity of $|B_\parallel|\sim 0$, giving a larger correlation field near
$|B_\parallel|\sim0$, than at larger parallel fields. This elongation,
demonstrating reduced flux sensitivity near $|B_\parallel|\sim0$, is
consistent with the $B_\parallel$ dependent effective mass and momentum
reversal symmetry breaking terms of our theory. Effects of nonplanarity
alone would result in a $B_\parallel$ independent correlation field. A
quantitative analysis is presented in Fig.~2.

The 2D slices in Fig.~1 also illustrate the fundamental symmetries of
conductance with respect to magnetic fields $B_\parallel$ and $B_\bot$:
when $B_\parallel=0$, conductance is symmetric under inversion of
$B_\bot$, $g(B_\bot)=g(-B_\bot)$ (see Fig.~1(b));  when $B_\bot=0$,
conductance is symmetric under inversion of $B_\parallel$,
$g(B_\parallel)=g(-B_\parallel)$ (see Fig.~1(c)).  When both $B_\parallel$
and $B_\bot$ are nonzero, the symmetry of conductance requires the
reversal of both fields, $g(B_\parallel,B_\bot)=g(-B_\parallel,-B_\bot)$
(see Fig.~1(a)) \cite{Buttiker}. The fact that a nonzero $B_\parallel$
breaks the symmetry $g(B_\bot)=g(-B_\bot)$ is a simple qualitative
demonstration that $B_\parallel$ breaks TRS \cite{Zumbuhl}. A quantitative
analysis of this effect is presented in Fig.~3.

To quantify the correlations of the various parameters used to generate
CF's --- including in particular $B_\parallel$ --- we define the
normalized correlation functions,
\begin{eqnarray}
C_v(B_\parallel)&=&\frac{\langle \delta g(B_\parallel,V) \delta
g(B_\parallel,V+v)\rangle}{\langle \delta g^2(B_\parallel)\rangle}\\
C_{b_\bot}(B_\parallel)&=&\frac{\langle \delta g(B_\parallel, B_\bot)
\delta g(B_\parallel, B_\bot+b_\bot)\rangle}{\langle \delta
g^2(B_\parallel)\rangle}\\
C_{b_{\parallel}}(B_{\parallel})&=&\frac{\langle \delta g(B_{\parallel}) \delta
g(B_{\parallel}+b_\parallel)\rangle}{\sqrt{\langle \delta
g^2(B_{\parallel})\rangle \langle \delta
g^2(B_{\parallel}+b_\parallel)\rangle}},
\end{eqnarray}
where $\langle \ldots \rangle$ is shorthand for $\langle \ldots
\rangle_{V,B_\bot}$, i.e., averaging over both gate voltage and $B_\bot$,
with $B_\bot$ sufficiently large to fully break TRS throughout the
measured range.

Theoretical expressions for the correlation functions in Eqs.~1--3 can be
found using the effective 2D Hamiltonian
\begin{equation}
\mathrm{\hat{H}}_{2D}=\frac{\mathbf{p}^{2}}{2m}-p_{\bot }^{2}\gamma
(B_{\Vert })+p_{\bot }^{3}\beta (B_{\Vert })+u(\mathbf{r}), \label{H2D}
\end{equation}
for electrons confined to a plane perpendicular to $\hat{z}$ \cite{Falko}.
Here, $\mathbf{p}=-i\hbar \nabla -\frac{e}{c}\mathbf{A}_{2D}$, with
$\mathrm{rot}\mathbf{A}_{2D}=B_{\bot }$, is the 2D momentum operator in
the plane, with component $p_{\bot }=\vec{p}\cdot \lbrack \vec{B}_{\Vert
}\times \vec{l}_{z}]/B_{\Vert }$ perpendicular to $B_\parallel$, and
$u(\mathbf{r})$ is the impurity and dot confining potential. The middle
terms in $\mathrm{\hat{H}}_{2D}$ arise from $p_\bot$-dependent subband
mixing: the $\gamma(B_\parallel)$ term lifts rotational symmetry with an
anisotropic mass enhancement \cite{2DEGaniso} but does not break TRS; the
$\beta(B_\parallel)$ arises from the broken spatial inversion symmetry of
the heterointerface and breaks TRS in presence of $B_\parallel$.

Extending the analysis of Ref.~\cite{Falko} to higher temperature yields a
correlation between conductance fluctuations at $B_{\parallel}$ and at
$B_{\parallel}+b_\parallel$ of the form
\begin{equation}
C_{b_\parallel}(B_{\parallel})= \left[ 1+
\frac{\tau_d^{-1}(B_{\parallel},b_{\parallel})}{\tau_{esc}^{-1}}
\right]^{\alpha},
\end{equation}
in the unitary ensemble, where $\tau_{esc}^{-1} = N\Delta/h$ is the escape
rate from the dot, with $\Delta=2\pi\hbar^2/m^*A$ the mean level spacing
of the corresponding closed dot (effective electron mass $m^*=0.067m_e$),
$\tau_d^{-1}$ is an additional escape rate due to orbital effects of
$B_\parallel$, as discussed below. The exponent $\alpha$ equals $-1$ in
the high temperature limit $kT\gg\ (\hbar \tau_{esc}^{-1}, \hbar
\tau_{d}^{-1}, \epsilon_Z)$, applicable in the present experiment, and
$-2$ in the low temperature limit, where $\epsilon_Z=g\mu_B B$ is the
Zeeman energy, with $g=-0.44$ for GaAs. The difference between the high
and low temperature regimes is caused by the necessity to average the
interference contributions coming from electrons at different energies.
For parallel fields with $\epsilon_Z \lesssim 3kT$, appropriate for the
present measurements, the deviation of Eq.~(5) from the full expression is
negligible \cite{FJLong}.

The additional escape rate $\tau_d^{-1}$ due to $B_\parallel$ is given by
\begin{eqnarray}
  \tau_d^{-1}(B_{\parallel},b_{\parallel}) &=& \frac{\tau
p_F^4}{8\hbar^2}\left[\gamma(B_{\parallel})-\gamma(B_{\parallel}+b_\parallel)\right]^2
\\\nonumber&+& \frac{\tau
p_F^6}{8\hbar^2}\left[\frac{\beta(B_{\parallel})-\beta(B_{\parallel}+b_\parallel)}{2}\right]^2
+ \frac{\zeta^2 p_F^2}{2\tau}\, b_\parallel^2,
\end{eqnarray}
where $\tau=\mu m^*/e$ is the elastic scattering time in a diffusive dot
or the crossing time $\tau=m^* L/p_F$ in a ballistic device, where $L$ is
the diameter of the device and $p_F=\hbar (2\pi n)^{1/2}$ is the Fermi
momentum. The $\zeta$ term describes effects of nonplanarity, including
interface roughness and dopant inhomogeneities, and also breaks TRS.

Writing the functions $\gamma(x)$ and $\beta(x)$ in Eq.~(6) in terms of
scale factors $\tilde\gamma$ and $\tilde\beta$ and normalized functions
$g(x)$ and $f(x)$ as $\gamma(x)=\tilde\gamma g(x)$ and
$\beta(x)=\tilde\beta f(x)$, we find $g(x)$ and $f(x)$ from
self-consistent simulations of the heterostructure \cite{SimDetails} and
treat $\tilde\gamma$ and $\tilde\beta$ as fit parameters. Below $\sim$ 2 T , the
normalized functions are well approximated by $g(x)\approx x^2$ and
$f(x)\approx x^3$ (see Fig.~2(b), insets); however, the full functions are
used for all comparison of theory and experiment. We note that
$\tilde\gamma$ and $\tilde\beta$ can also be obtained from the
heterostructure simulations, giving values in reasonable agreement with
those obtained from the fits.

\begin{figure}[t]
              \label{fig2ucfcorr}
              \includegraphics[width=3in]{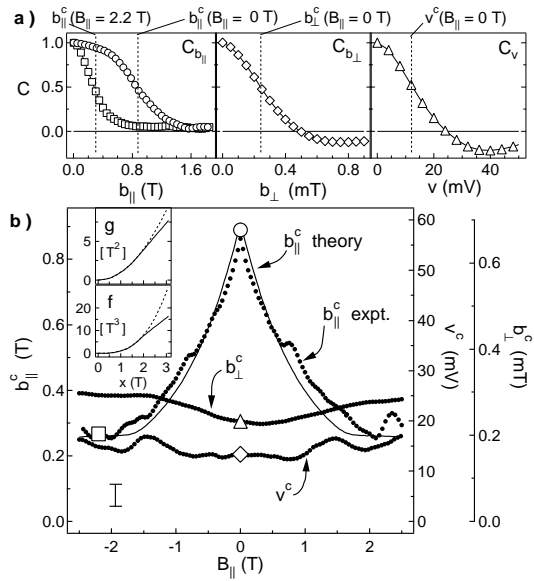}
      \caption{\footnotesize{a) Correlation functions
      $C_{b_\parallel}$ at $B_\parallel=0,\, 2.2\, \mathrm{T}$
      (open circles, squares) as well as $C_{b_\bot}$ (open
      diamonds) and $C_v$ (open triangles) at $B_\parallel=0$.
      Half width at half maximum values give the characteristic
      voltage $v^c$ and fields $b_\bot^c$ and $b_\parallel^c$,
      shown in b) (solid circles) as a function of $B_\parallel$.
      Markers in b) refer to corresponding curves in a).
      The solid curve shows the three-parameter theory.
      A typical error bar is indicated.
      Insets: $g(x)$ and $f(x)$ used for fits (see text)
      as obtained from numerical simulations (solid curves)
      as well as quadratic and cubic low field
      approx. (dashed curves).
      }}
\end{figure}

Figure~2(a) shows experimental correlation functions, $C_{b_\parallel}$,
$C_{b_\bot}$, and $C_v$, for representative parallel fields, as indicated.
The corresponding characteristic voltage $v^c$ and fields $b_\parallel^c$
and $b_\bot^c$ are shown in Fig.~2(b) as a function of $B_\parallel$, as
obtained from the half width at half maximum (HWHM) values of the
correlation functions, indicated by dashed lines in Fig.~2(a). It is
evident from Fig.~2(b) that both $b^c_\bot$ and $v^c$ {\em are independent
of $B_\parallel$} within the error bars, in agreement with theory and
previous experiments \cite{Folk}. (An alternative procedure, not shown,
for extracting these same quantities from the slopes of log-power spectra
of CF's yields similar values for $v^c$ and $b_\parallel^c$  that are
again independent of $B_\parallel$, within error bars.)

In contrast, the parallel field correlation length, $b^c_\parallel$, shown
in Fig.~2(b) decreases substantially from its zero-field value on a field
scale of $\sim 1$\,T. Good agreement with theory is found: the solid curve
in Fig.~2(b) is the best-fit (described below) theoretical HWHM
correlation field for $C_{b_\parallel}(B_\parallel)$ obtained from
Eq.~(5). This decrease is due to the $\gamma$ and $\beta$ terms in Eq. (6)
and cannot be accounted for with the $\zeta$ term alone.

\begin{figure}[t]
              \label{fig3ucfcorr}
          \includegraphics[width=3in]{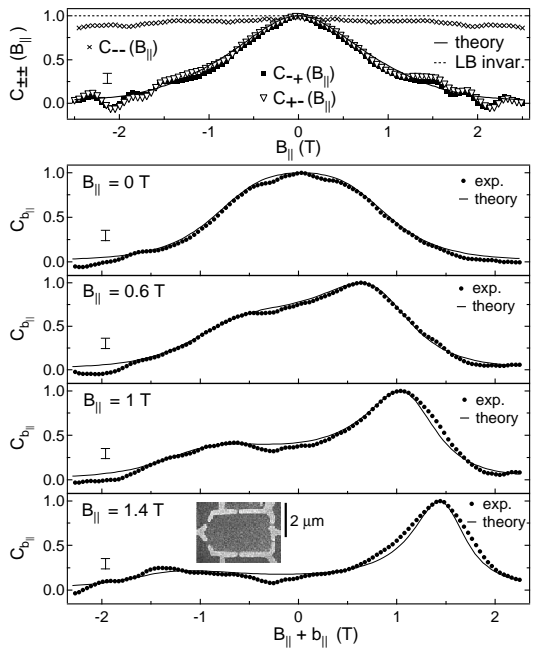}
        \caption{\footnotesize{Top: Cross-correlations of CF's at
        ($B_{\parallel}$,$B_\bot$) with CF's at
        ($-B_{\parallel}$,$-B_\bot$) ($C_{--}$, crosses),
        ($-B_{\parallel}$,$B_\bot$) ($C_{-+}$, squares)
        and ($B_{\parallel}$,$-B_\bot$) ($C_{+-}$, triangles)
        as a function of $B_\parallel$. Landauer-B\"uttiker (LB) symmetry for full field reversal gives  $C_{--} = 1$ (Dashed line). Below: Correlation
        functions $C_{b_\parallel}(B_\parallel)$ at
        $B_\parallel=0,\, 0.6,\, 1,\, 1.4\, \mathrm{T}$.
        Solid curves show theory based on Eq.\,(5). Typical error bars as indicated.
       }}
\end{figure}

Symmetries of conductance in parallel and perpendicular fields are
investigated in Fig.~3.  We define the cross-correlation functions
\begin{equation}
C_{\pm\pm}(B_\parallel)=\frac{\langle \delta g(B_\parallel,B_\bot)
\delta g(\pm B_\parallel,\pm
B_\bot)\rangle}{\sqrt{\langle \delta g^2(B_\parallel,B_\bot)\rangle
\langle \delta g^2(\pm B_\parallel,\pm
B_\bot)\rangle}}.
\end{equation}
With this definition, the first (second) subscript index of $C$ refers to whether $B_\parallel (B_\bot)$ is reversed when computing the correlation function.
Correlations for total field reversal, $C_{--}$ (i.e., both $B_\parallel$
and $B_\bot$ inverted) remain near unity for all parallel
fields, as expected from the full Landauer-B\"uttiker (Onsager)
symmetry (see Fig.~3(a)).  Deviations from a perfect correlation $C_{--}=1$
are small, indicating that the confining potential of the dot did not
drift significantly over periods of a day. Figure~3(a) also shows
conductance fluctuations at ($B_\parallel$,$B_\bot$) and
($B_\parallel$,$-B_\bot$) ($C_{+-}$) become uncorrelated ($C_{+-}\sim0$)
at parallel fields of a few tesla, indicating the field scale at which
$B_\parallel$ breaks TRS. Within error bars, $C_{-+}$ is indistinguishable
from $C_{+-}$, as expected from Landauer-B\"uttiker symmetry, $\delta
g(-B_\parallel,B_\bot)=\delta g(B_\parallel, -B_\bot)$. The theoretical
cross-correlation using Eq.~(5), shown as a solid curve in Fig.~3(a), is in
very good agreement with experiment data.

Finally, we discuss the full correlation, $C_{b_\parallel}(B_\parallel)$,
of CF's at $B_\parallel$ with CF's at $B_\parallel+b_\parallel$.
Representative curves for $B_{\parallel}=0,\, 0.6,\, 1,\, 1.4\,
\mathrm{T}$ are shown in the lower part of Fig.~3 as a function of
$B_\parallel+b_\parallel$, along with best-fit theory curves based on
Eq.~(5).  Besides the perfect correlation at $b_\parallel=0$
($C_{b_\parallel=0}=1$), there is an ``echo" of correlations, both in
experiment and theory, that occurs at $b_{\parallel} \sim
-2B_{\parallel}$.  Within the present theory, this field-reversed
correlation ``echo", $C_{-2B_\parallel}$, is suppressed from unity only to
the extent that parallel field breaks TRS. The agreement between theory
and experiment, including the unusual asymmetric curves in Fig.~3, is
quite good.  A single, consistent set of three parameters ($\tilde\gamma$,
$\tilde\beta$, $\zeta$) have been obtained from fits of Eq.~(5) to $131$
curves like those in Fig.~3, ranging over $-2.5\, \mathrm{T}\leq
B_\parallel\leq 4\, \mathrm{T}$. We emphasize that {\em all} theory curves
shown in Fig.~2 and Fig.~3 used this single set of three fit parameters
and were not individually fit.  The values obtained in this way were
$\tilde\gamma=11\pm2\, \times10^{-4}\, [m^*]^{-1}T^{-2}$,
$\tilde\beta=4\pm4\, \times 10^{-4}[m^*p_F]^{-1}T^{-3}$, and
$\zeta=44\pm8\, \times 10^{-3}[p_F]^{-1}T^{-1}$, consistent within the
error bars with values extracted for the $3\, \mathrm{\mu m^2}$ dot.  We
note that parameters $\tilde\beta$ and $\zeta$ obtained from the
parallel-field-induced crossover from the orthogonal to the unitary
ensemble \cite{Zumbuhl} are in good agreement. The self-consistent
simulations give theoretical values of $\tilde\gamma=35\times10^{-4}\,
[m^*]^{-1}T^{-2}$, $\tilde\beta=3\times10^{-4}[m^*p_F]^{-1}T^{-3}$.

In summary, orbital effects of an in-plane magnetic field $B_\parallel$
were experimentally investigated using the high sensitivity of CF's to
magnetic flux in a large quasiballistic quantum dot. Detailed quantitative
comparison of correlations of CF's induced by $B_\parallel$ with theory
developed here reveal the mechanisms of coupling, including an induced
anisotropic effective mass, the breaking of time-reversal symmetry due to
the heterostructure asymmetry and effects of nonplanarity. In the present
experiment, spin-orbit coupling is weak. On the other hand, the combined
influence of stronger spin-orbit coupling \cite{Zumbuhl} and parallel
fields is expected to yield interesting additional features in the
correlations and symmetries of CF's \cite{JanhPietSo}.  These remain to be
investigated experimentally.

We thank I. Aleiner, P. Brouwer, J. Folk and D. Goldhaber-Gordon for useful discussions. This
work was supported in part by DARPA-QuIST, DARPA-SpinS, ARO-MURI and
NSF-NSEC. Work at UCSB was supported by QUEST, an NSF Science and
Technology Center. JBM acknowledges partial support from NDSEG, and VIF from
the Royal Society end EPSRC.


{

\small \vspace{-0.2in}
\begin{thebibliography}{10}

\expandafter\ifx\csname bibnamefont\endcsname\relax
                \def\bibnamefont#1{#1}\fi
\expandafter\ifx\csname bibfnamefont\endcsname\relax
                \def\bibfnamefont#1{#1}\fi
\expandafter\ifx\csname url\endcsname\relax
                \def\url#1{\texttt{#1}}\fi
\expandafter\ifx\csname urlprefix\endcsname\relax\def\urlprefix{URL }\fi
\expandafter\ifx\csname bibinfo\endcsname\relax \def\bibinfo#1#2{#2}\fi
\expandafter\ifx\csname eprint\endcsname\relax \def\eprint#1{#1}\fi
\small

\bibitem{Falko}
\bibinfo{author}{\bibfnamefont{V.~I.} \bibnamefont{Fal'ko}} \bibnamefont{and}
        \bibinfo{author}{\bibfnamefont{T.} \bibnamefont{Jungwirth}},
               \bibinfo{journal}{Phys. Rev. B}
               \textbf{\bibinfo{volume}{65}}, \bibinfo{pages}{81306}
               (\bibinfo{year}{2002}).

\bibitem{Meyer}
\bibinfo{author}{\bibfnamefont{J.~S.} \bibnamefont{Meyer}},
\bibinfo{author}{\bibfnamefont{A.} \bibnamefont{Altland}} \bibnamefont{and}
\bibinfo{author}{\bibfnamefont{B.~L.} \bibnamefont{Altshuler}},
               \bibinfo{journal}{Phys. Rev. Lett.}
               \textbf{\bibinfo{volume}{89}}, \bibinfo{pages}{206601}
               (\bibinfo{year}{2002});
               \bibinfo{author}{\bibfnamefont{J.~S.} \bibnamefont{Meyer}},
               \bibinfo{author}{\bibfnamefont{V.~I.}
               \bibnamefont{Falko}} \bibnamefont{and}
               \bibinfo{author}{\bibfnamefont{B.~L.} \bibnamefont{Altshuler}},
               \bibnamefont{in NATO Science Series II, Vol. 72, edited
                by I.V. Lerner, B.L. Altshuler, V.I. Fal'ko and
                T. Giamarchi (Kluwer Academic Publishers,
                Dordrecht, 2002), pp. 117-164.}

\bibitem{Zumbuhl}
\bibinfo{author}{\bibfnamefont{D.~M.} \bibnamefont{Zumb\"uhl}},
        \bibinfo{author}{\bibfnamefont{J.~B.} \bibnamefont{Miller}},
        \bibinfo{author}{\bibfnamefont{C.~M.} \bibnamefont{Marcus}},
        \bibinfo{author}{\bibfnamefont{K.} \bibnamefont{Campman}},
        \bibnamefont{and}
        \bibinfo{author}{\bibfnamefont{A.~C.} \bibnamefont{Gossard}},
        \bibinfo{journal}{Phys. Rev. Lett.}
        \textbf{\bibinfo{volume}{89}}, \bibinfo{pages}{276803}
        (\bibinfo{year}{2002});
\bibinfo{author}{\bibfnamefont{D.~M.} \bibnamefont{Zumb\"uhl}} \emph{et~al.},
        \bibinfo{journal}{(to be published)}.

\bibitem{Metals}
\bibinfo{author}{\bibfnamefont{Z.} \bibnamefont{Ovadyahu}} \emph{et al.},
                \bibinfo{journal}{Phys. Rev. B}
                \textbf{\bibinfo{volume}{32}}, \bibinfo{pages}{781}
                (\bibinfo{year}{1985});
\bibinfo{author}{\bibfnamefont{N.} \bibnamefont{Giordano}} \bibnamefont{and}
         \bibinfo{author}{\bibfnamefont{M.~A.} \bibnamefont{Pennington}},
                \bibinfo{journal}{Phys. Rev. B}
                \textbf{\bibinfo{volume}{47}}, \bibinfo{pages}{9693}
                (\bibinfo{year}{1993}).

\bibitem{SiMOSFET}
\bibinfo{author}{\bibfnamefont{D.~J.} \bibnamefont{Bishop}} \emph{et al.},
                \bibinfo{journal}{Phys. Rev. B}
                \textbf{\bibinfo{volume}{26}}, \bibinfo{pages}{773}
                (\bibinfo{year}{1982});
\bibinfo{author}{\bibfnamefont{P.~M.} \bibnamefont{Mensz}} \bibnamefont{and}
         \bibinfo{author}{\bibfnamefont{R.~G.} \bibnamefont{Wheeler}},
                \bibinfo{journal}{Phys. Rev. B}
                \textbf{\bibinfo{volume}{35}}, \bibinfo{pages}{2844}
                (\bibinfo{year}{1987});
\bibinfo{author}{\bibfnamefont{U.} \bibnamefont{Kunze}},
                \bibinfo{journal}{Phys. Rev. B}
                \textbf{\bibinfo{volume}{35}}, \bibinfo{pages}{9168}
                (\bibinfo{year}{1987}).

\bibitem{2DEG}
\bibinfo{author}{\bibfnamefont{B.~J.~F.} \bibnamefont{Lin}} \emph{et al.},
                \bibinfo{journal}{Phys. Rev. B}
                \textbf{\bibinfo{volume}{29}}, \bibinfo{pages}{927}
                (\bibinfo{year}{1984}).

\bibitem{2DEGaniso}
\bibinfo{author}{\bibfnamefont{J.~M.} \bibnamefont{Heisz}} \bibnamefont{and}
         \bibinfo{author}{\bibfnamefont{E.} \bibnamefont{Zaremba}},
                \bibinfo{journal}{Phys. Rev. B}
                \textbf{\bibinfo{volume}{53}}, \bibinfo{pages}{13594}
                (\bibinfo{year}{1996}).

\bibitem{Focus}
\bibinfo{author}{\bibfnamefont{K.} \bibnamefont{Ohtsuka}} \emph{et al.},
                \bibinfo{journal}{Physica B}
                \textbf{\bibinfo{volume}{249-251}}, \bibinfo{pages}{780}
                (\bibinfo{year}{1998});
\bibinfo{author}{\bibfnamefont{K.} \bibnamefont{Oto}} \emph{et al.},
                \bibinfo{journal}{Physica E}
                \textbf{\bibinfo{volume}{11}}, \bibinfo{pages}{177}
                (\bibinfo{year}{2001}).

\bibitem{Schlesinger}
\bibinfo{author}{\bibfnamefont{Z.} \bibnamefont{Schlesinger}} \emph{et al.},
               \bibinfo{journal}{Phys. Rev. Lett.}
               \textbf{\bibinfo{volume}{50}}, \bibinfo{pages}{2098}
               (\bibinfo{year}{1983}).

\bibitem{MR}
\bibinfo{author}{\bibfnamefont{D.~R.} \bibnamefont{Leadley}} \emph{et al.},
               \bibinfo{journal}{Semicond. Sci. Technol.}
               \textbf{\bibinfo{volume}{5}}, \bibinfo{pages}{1081}
               (\bibinfo{year}{1990});
\bibinfo{author}{\bibfnamefont{J.~J.} \bibnamefont{Mares}} \emph{et al.},
               \bibinfo{journal}{Phys. Rev. Lett.}
               \textbf{\bibinfo{volume}{80}}, \bibinfo{pages}{4020}
               (\bibinfo{year}{1998});
\bibinfo{author}{\bibfnamefont{O.~N.} \bibnamefont{Makarovskii}} \emph{et al.},
               \bibinfo{journal}{Phys. Rev. B}
               \textbf{\bibinfo{volume}{62}}, \bibinfo{pages}{10908}
               (\bibinfo{year}{2000});
\bibinfo{author}{\bibfnamefont{P.} \bibnamefont{Svoboda}} \emph{et al.},
               \bibinfo{journal}{Physica E}
               \textbf{\bibinfo{volume}{12}}, \bibinfo{pages}{315}
               (\bibinfo{year}{2002}).

\bibitem{Choi}
\bibinfo{author}{\bibfnamefont{K.~K.} \bibnamefont{Choi}} \emph{et. al.},
               \bibinfo{journal}{Phys. Rev. B}
               \textbf{\bibinfo{volume}{38}}, \bibinfo{pages}{12362}
               (\bibinfo{year}{1988}).

\bibitem{Reynolds}
\bibinfo{author}{\bibfnamefont{D.~C.} \bibnamefont{Reynolds}} \emph{et. al.},
               \bibinfo{journal}{Phys. Rev. B}
               \textbf{\bibinfo{volume}{35}}, \bibinfo{pages}{4515}
               (\bibinfo{year}{1987});
\bibinfo{author}{\bibfnamefont{V.} \bibnamefont{Kirpichev}} \emph{et. al.},
               \bibinfo{journal}{JETP Lett.}
               \textbf{\bibinfo{volume}{51}}, \bibinfo{pages}{436}
               (\bibinfo{year}{1990});
\bibinfo{author}{\bibfnamefont{I.} \bibnamefont{Kukushkin}} \emph{et. al.},
               \bibinfo{journal}{JETP Lett.}
               \textbf{\bibinfo{volume}{53}}, \bibinfo{pages}{334}
               (\bibinfo{year}{1991}).

\bibitem{MarcusUCF}
\bibinfo{author}{\bibfnamefont{C.~M.} \bibnamefont{Marcus}} \emph{et~al.},
               \bibinfo{journal}{Phys. Rev. Lett.}
               \textbf{\bibinfo{volume}{69}}, \bibinfo{pages}{506}
               (\bibinfo{year}{1992});
\bibinfo{author}{\bibfnamefont{I.~H.} \bibnamefont{Chan}} \emph{et~al.},
               \bibinfo{journal}{Phys. Rev. Lett.}
               \textbf{\bibinfo{volume}{74}}, \bibinfo{pages}{3876}
               (\bibinfo{year}{1995});
\bibinfo{author}{\bibfnamefont{A.~G.} \bibnamefont{Huibers}} \emph{et~al.},
               \bibinfo{journal}{Phys. Rev. Lett.}
               \textbf{\bibinfo{volume}{81}}, \bibinfo{pages}{1917}
               (\bibinfo{year}{1998}).

\bibitem{Folk}
\bibinfo{author}{\bibfnamefont{J.~A.} \bibnamefont{Folk}} \emph{et~al.},
               \bibinfo{journal}{Phys. Rev. Lett.}
               \textbf{\bibinfo{volume}{86}}, \bibinfo{pages}{2102}
               (\bibinfo{year}{2001}).

\bibitem{Buttiker}
\bibinfo{author}{\bibfnamefont{M.} \bibnamefont{B\"uttiker}} and
        \bibinfo{author}{\bibfnamefont{Y.} \bibnamefont{Imry}},
               \bibinfo{journal}{J. Phys. C.}
               \textbf{\bibinfo{volume}{L}}, \bibinfo{pages}{467}
               (\bibinfo{year}{1985}).

\bibitem{FJLong}
\bibinfo{author}{\bibfnamefont{V.~I.} \bibnamefont{Fal'ko}} and
        \bibinfo{author}{\bibfnamefont{T.} \bibnamefont{Jungwirth}},
               \bibinfo{journal}{unpublished.}

\bibitem{SimDetails}
\bibnamefont{The simulations take into account band offsets, Hartree and exchange-correlation potentials as well as the
potential due to residual acceptor impurities, as obtained from the onset
of second subband population.}

\bibitem{JanhPietSo}
\bibinfo{author}{\bibfnamefont{J.~H.} \bibnamefont{Cremers}},
        \bibinfo{author}{\bibfnamefont{P.~W.} \bibnamefont{Brouwer}} and
        \bibinfo{author}{\bibfnamefont{V.~I.} \bibnamefont{Fal'ko}},
               \bibfnamefont{cond-mat/0304222;}
\bibinfo{author}{\bibfnamefont{I.} \bibnamefont{Aleiner}} and
        \bibinfo{author}{\bibfnamefont{V.~I.} \bibnamefont{Fal'ko}},
               \bibinfo{journal}{Phys. Rev. Lett. }
               \textbf{\bibinfo{volume}{87}}, \bibinfo{pages}{256801}
               (\bibinfo{year}{2001}).

\end{thebibliography}
 }
\end{document}